\documentclass[fleqn,10pt]{wlscirep}
\usepackage[utf8]{inputenc}
\usepackage[T1]{fontenc}
\usepackage{float} 
\usepackage{siunitx}
\DeclareSIUnit\gauss{G}
\title{Different sensitivities of two optical magnetometers realized in the same experimental arrangement.}

\author[1,*]{Piotr Put}
\author[1]{Kacper Popiołek}
\author[1]{Szymon Pustelny}
\affil[1]{Marian Smoluchowski Institute of Physics, Jagiellonian University, \L ojasiewicza 11, 30-348 Krak\'ow, Poland}

\affil[*]{piotr.put@doctoral.uj.edu.pl}

\begin{abstract}
In this article, operation of optical magnetometers detecting static (DC) and oscillating (AC) magnetic fields is studied and comparison of the devices is performed. To facilitate the comparison, the analysis is carried out in the same experimental setup, exploiting nonlinear magneto-optical rotation. In such a system, a control over static-field magnitude or oscillating-field frequency provides detection of strength of the DC or AC fields. Polarization rotation is investigated for various light intensities and AC-field amplitudes, which allows to determine optimum sensitivity to both fields. With the results, we demonstrate that under optimal conditions the AC magnetometer is about ten times more sensitive than its DC counterpart, which originates from different response of the atoms to the fields. Bandwidth of the magnetometers is also analyzed, revealing its different dependence on the light power. Particularly, we demonstrate that bandwidth of the AC magnetometer can be significantly increased without strong deterioration of the magnetometer sensitivity. This behavior, combined with the ability to tune the resonance frequency of the AC magnetometer, provide means for ultra-sensitive measurements of the AC field in a broad but spectrally-limited range, where detrimental role of static-field instability is significantly reduced.
\end{abstract}
\begin{document}

\flushbottom
\maketitle

\thispagestyle{empty}

\section*{Introduction}

Over the past 15 years, research on magneto-optical phenomena led to the development of various types of optical magnetometers \cite{magnetometrybook}. This development led to significantly enhanced capabilities of all magnetic-field sensors. For example, the most sensitive magnetometer ever constructed is the so-called spin-exchange relaxation-free (optical) magnetometer, which reached a near-DC sensitivity of \SI{1.6}{\pico\gauss\per\sqrt{\hertz}} \cite{dang2010ultrahigh}. Besides sensitivity, other advantages of optical magnetometers are: low price and maintenance, low power consumption and relative ease of miniaturization. This makes the devices attractive for applications ranging from fundamental research (searches for 5th force, dark matter and dark energy, electric dipole moment, Lorentz and CPT violation etc.)\cite{pustelny2013global,safronova2018search,smiciklas2011new,kornack2005test,vasilakis2009limits}, through semi-practical or emerging applications (zero- and ultra-low-field nuclear magnetic resonance and magnetic resonance imaging, archeology, palomagnetism)\cite{ledbetter2011near,xu2006magnetic,romalis2011atomic,geomagnetometry}, to purely utilitarian applications (medical diagnostics, prospecting for natural resources, military surveys)\cite{shah2013compact, schultz2016integration}.

Modern optical magnetometers can be used for detection of (quasi-)static (DC) and oscillating (AC) magnetic fields (in a narrow tunable frequency range). As shown in the literature (see, for example, Ref.~\cite{higbie2006robust,pustelny2008magnetometry,Chalupczak} and references therein), the sensitivity of DC and AC magnetometers can vary significantly despite the fact the fields are measured in similar or even identical experimental arrangements. One may associate this difference with 1/$f$ noise, which affects DC measurements more strongly, but a true origin of the difference is of fundamental nature, stemming from a resonant response of the medium to the AC magnetic field. Specifically, under the resonance condition, when the frequency of the oscillating field coincides with the spin precession frequency (the Larmor frequency), an effect of even a small perturbation can accumulate, leading to significant modifications of properties of a whole medium.

In this article, we analyze the performance of DC and AC optical magnetometers. The measurements are carried out in the same experimental arrangement. This goal is achieved by application of radio-frequency nonlinear magneto-optical rotation (NMOR)\cite{NMOR,ledbetter2007detection}, i.e., rotation of the polarization plane of linearly polarized, resonant light, propagating through a medium subjected to static and oscillating magnetic fields \cite{Savukov;2005}. In such an arrangement, the strongest polarization rotation is observed when the frequency of the oscillating field is tuned to the frequency splitting of adjacent Zeeman sublevels (the Larmor frequency). A control over the static-field magnitude or the oscillating-field frequency allows for the detection of either AC or DC fields. Particularly, in the case of static-field measurements, determination of the resonance frequency, corresponding to maximum rotation of the polarization, provides information about the static-field strength. Alternatively, by tuning the static field so that the Zeeman splitting coincides with the AC-field frequency and analyzing the amplitude of the rotation signal we are able to measure the strength of the oscillating field. 

Herein, we investigate such parameters of the magnetometers as sensitivity and bandwidth, analyzing the quantities as functions of the light power and AC-field amplitude. Determination of dependences allow us to optimize the performance of the devices.

\section*{Materials and Methods}
\subsection*{Experimental arrangement}
\begin{figure}[H]
\centering
\includegraphics[width=0.375\linewidth]{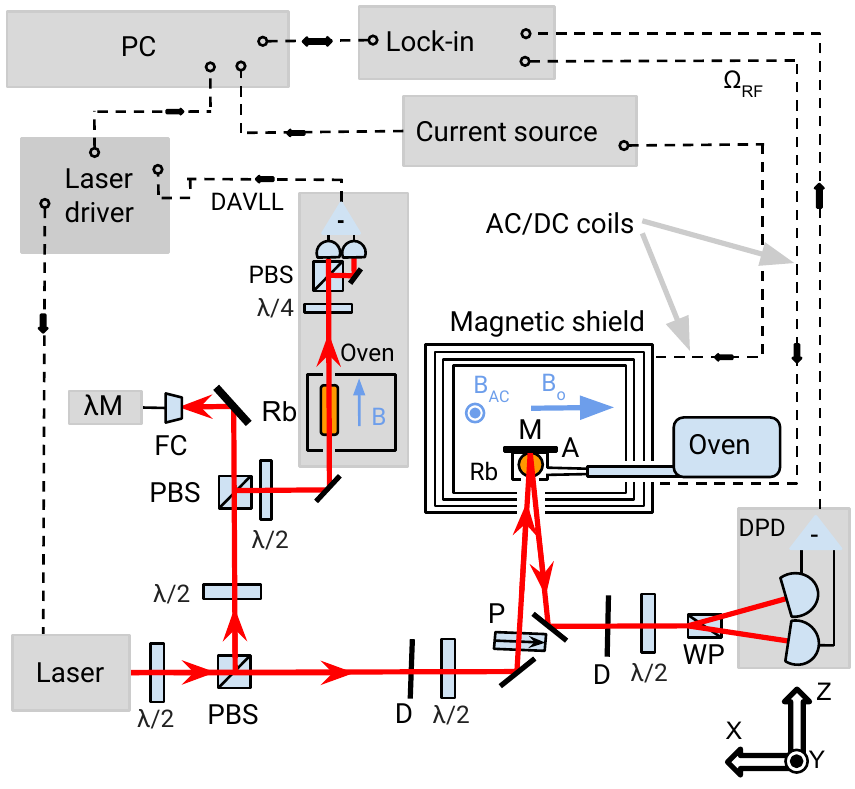}
\caption {Diagram of the experimental apparatus used for magnetic-field detection. $\lambda$/2 is a half-wave plate, $\lambda$/4 is a quarter-wave plate, PBS is a polarizating beam splitter, FC is a fiber-optic coupler, $\lambda$M is a wavelength meter, Rb is a rubidium vapor cell, D is a diaphragm, P is a polarizer, M is a mirror, A is an inlet of hot air used for cell heating, WP is a Wollaston prism, DPD is a differential photodiode, PC is a computer, DAVLL is a dichroic atomic vapor laser lock system, $B$ is a static field for the DAVLL system, $B_{AC}$ is an oscillating field and $B_0$ is a static field used for magnetometric purposes. Directions of magnetic fields are marked by bright blue arrows.}
\label{fig:Setup}
\end{figure}

Figure~\ref{fig:Setup} presents an experimental setup used in our measurements. A diode laser (Toptica DL100) generates light detuned by about $-$480 MHz from the $F=2\rightarrow F'=1$ transition of the rubidium $D_1$ line (\SI{795}{\nano\meter}). Right after the laser, the light is split off and its fraction is sent to a side arm for wavelength monitoring with a wavemeter (HighFinesse/Angstrom Wavelength meter WS-U) and wavelength stabilization with a dichroic atomic vapor laser lock system\cite{DAVLL}. The main beam, of 4 mm in diameter, is used to illuminate a spherical (\SI{4}{\centi\meter} in diameter) atomic vapor cell, which walls are coated with an antirelaxation layer (paraffin). A narrow capillary, attached to the side of the cell, contains a metallic droplet of isotopically enriched $^{87}$Rb, serving as an atomic reservoir. The cell is placed inside a cylindrical magnetic shield, consisting of four mu-metal layers (Twinleaf MS-1). While the shield significantly reduces external, uncontrollable magnetic fields (a shielding factor on the order of 10$^{6}$), a set of magnetic-field coils mounted inside is used to compensate for residual fields (including first-order gradients) and to generate a static field along $x$ (typically on the order of \SI{10}{\milli\gauss}). The coils are powered with a precision current source (Keithley 6220). An additional oscillating field is generated along $y$ and its amplitude and frequency is controlled with a generator built in a lock-in amplifier (Zurich Instruments HF2LI). The rubidium-vapor cell is heated with hot air to about \SI{50}{\degreeCelsius} (resulting in concentration of $3\times 10^{11}$ atoms/cm$^3$). The probe light is $x$-polarized and its intensity is controlled between 0 to \SI{1000}{\micro\watt} with a half-wave plate and a high-quality crystal polarizer. After passing through the cell, the light is reflected and leaves the shield throughout the same shield's optical port at a small angle with respect to the incident beam. This increases light-atom interaction length and hence results in an improvement of an NMOR signal. The polarization rotation of the light is measured with a balanced polarimeter, consisting of a Wollaston prism and a balanced photodetector (Thorlabs PDB450A). The photodetector difference signal is demodulated with the lock-in amplifier at the first harmonic of the AC-field frequency. The whole system is controlled with a computer, which also provides data storage.

\subsection*{Methods}

In NMOR exploiting oscillating magnetic field, maximum polarization rotation is observed when the AC-field frequency $\omega_{AC}$ coincides with the Larmor frequency $\omega_L$ ($\omega_L=g\mu_BB_0/\hbar$, where $B_0$ is the field strength, $g$ is the Land\'e factor, $\mu_B$ is the Bohr magneton, and $\hbar$ is the reduced Planck constant) (for simplicity only the linear Zeeman effect is considered). The rotation signal, measured versus the AC-field frequency, has a resonant character and the width of the resonance is determined by a ground-state relaxation rate. The $B_0$-field change results in modification of the observed resonance condition and hence change of the rotation signal. For the DC-field measurement, it is convenient to detect the rotation component phased with the oscillating field (an in-phase component of the signal), which is dispersively shaped [see Fig.~\ref{fieldchange}(a)] thus provides information about about the amplitude and sign of the change.

Changes of the rotation signal are also observed when the amplitude of the oscillating field $B_{AC}$ is modified, while other parameters, including the static field $B_0$, are constant. In this case, however, the resonance position remains the same but its amplitude increases. If calibrated, the dependence of the signal on the AC-field amplitude enables quantitative detection of the AC-field amplitude $B_{AC}$.

The dependence of the NMOR-signal amplitude and in-phase component on the static-field strength $B_0$ and the AC-field amplitude $B_{AC}$ enables detection of the DC and AC magnetic fields. To compare the DC- and AC-field measurements, we analyze the NMOR signals versus the light power and AC-field amplitude. By measuring the amplitude and width of the NMOR resonances, we are able to determine a slope of the in-phase component at resonance and hence determine the response of the magnetometer to DC-field changes. Similarly, by analyzing the amplitude of the NMOR signal for different AC-field amplitudes, we are able to determine the response of the magnetometer to oscillating fields. These dependences, after noise normalization,  yield information about sensitivity of the measurements. Finally, we determine the bandwidth of the DC magnetometer by modulating the static field and bandwidth of the AC magnetometer by applying a square-wave pulse of the oscillating field.

\section*{Results and Discussion}
\subsection*{System response to the DC and AC field stepwise change}

\begin{figure}[H]
\centering
\includegraphics[width=0.5625\linewidth]{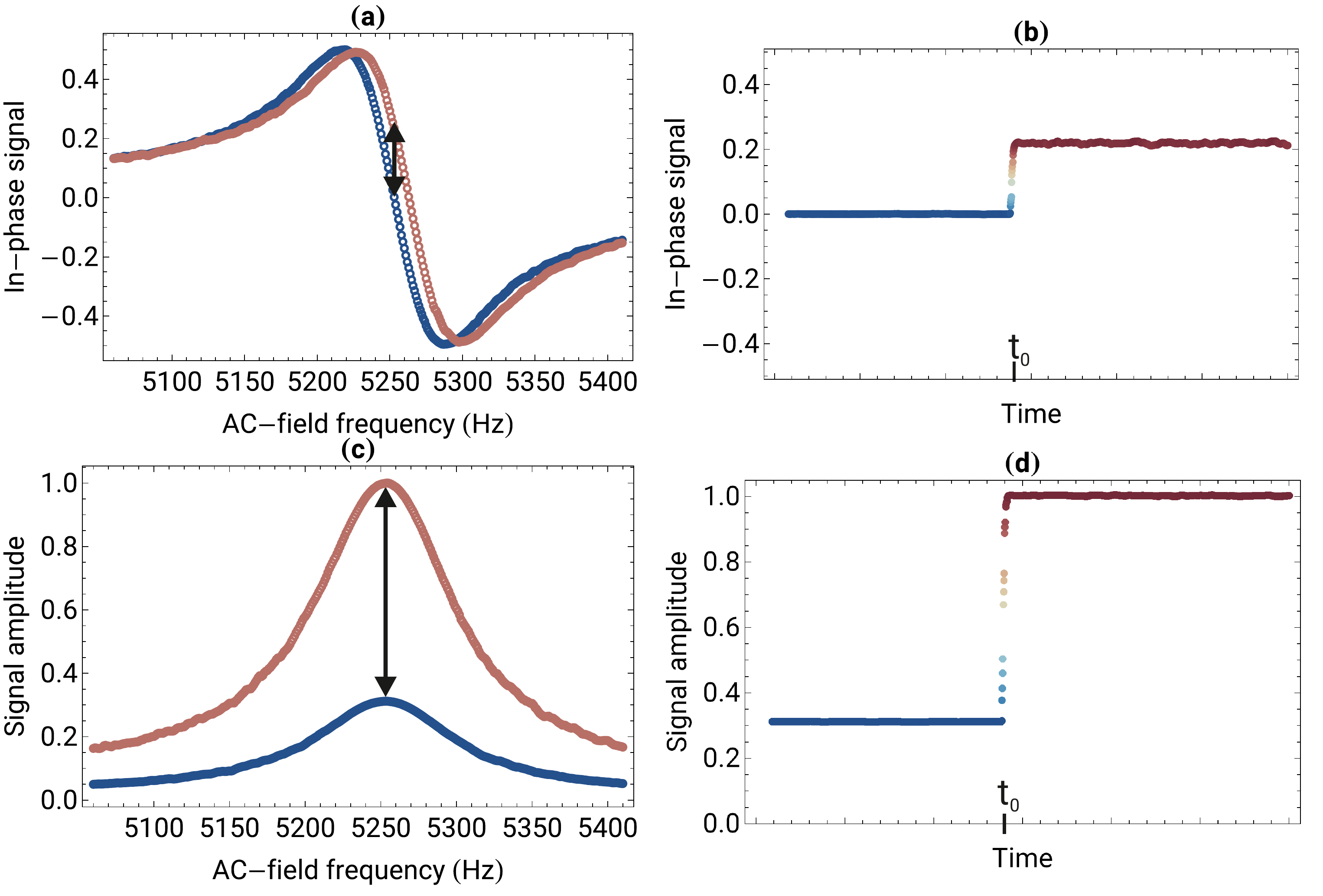}
\caption{Nonlinear magneto-optical rotation measured with the DC and AC magnetometers. (a) In-phase component of the signal recorded versus the AC-field frequency near resonance ($\omega_{AC}\approx\omega_L \approx$ \SI{5.25}{\kilo\hertz}) for two static-field values. The change of the DC field by \SI{15}{\micro\gauss} leads to the 10-Hz resonance shift. (b) Change in the in-phase component of the signal for the fixed frequency (initially $\omega_{AC}=\omega_L\approx\SI{5.25}{\kilo\hertz}$) when the static field is modified by \SI{15}{\micro\gauss} at $t=t_0$. (c) Amplitude of the NMOR signal measured for two different AC-field amplitudes and a fixed static field of $\omega_L\approx \SI{5.25}{\kilo\hertz}$.  (d) Rotation signal measured for $\omega_{AC}=\omega_L\approx\SI{5.25}{\kilo\hertz}$ when amplitude of the AC field is changed by \SI{15}{\micro\gauss} at $t=t_0$ and other experimental parameters remain unchanged. The vertical arrows in (a) and (c) mark the NMOR signals at $\omega_{AC}=\SI{5.25}{\kilo\hertz}$ when DC and AC fields are changed by \SI{15}{\micro\gauss} [signals presented in (b) and (d)].}
\label{fieldchange}
\end{figure}

To analyze the response of the DC and AC magnetometers to the field changes, we measure both the in-phase component and amplitude of the polarization rotation. As discussed above, the change of static magnetic field leads to the shift of the observed resonance. To demonstrate this behavior, Fig.~\ref{fieldchange}(a) presents the NMOR signals measured for two static magnetic fields. As shown, a change of a static field of \SI{7.5}{\milli\gauss} by \SI{15}{\micro\gauss} results in a shift of the resonance position by about \SI{10}{\hertz}. Similarly, Fig.~\ref{fieldchange}(c) presents the NMOR signal observed for two different AC-field amplitudes. In this case, however, the change of the field does not lead to the shift of the resonance but rather change of its amplitude. In particular, when an AC-field amplitude of \SI{1}{\micro\gauss}$_{rms}$ is increased by \SI{15}{\micro\gauss}, the NMOR-signal amplitude changes from about 0.25 to 1.  

The difference between the DC- and AC-magnetometer response manifests more strongly when the time traces of the signals are analyzed [Figs.~\ref{fieldchange}(b) \& (d)]. Initially, both magnetometers operate under the resonance condition ($\omega_{AC}=\omega_L$) when maximum rotation but zero of the in-phase component of the NMOR signal are recorded. At time $t_0$, the static [Fig.~\ref{fieldchange}(b)] or oscillating [Fig.~\ref{fieldchange}(d)] magnetic field is abruptly changed by \SI{15}{\micro\gauss}. The changes result in modification of the NMOR signals, which, after some transient time, determined by the magnetometers' bandwidths, settle at new values (It is noteworthy that under special conditions, a transient signals not limited by the magnetometer bandwidth may be observed \cite{grewal2018transient}. While these signals may be used for detection of fast changes of the field, their magnetometric application ranges beyond the scope of the paper.). The two plots clearly show that even the same change in the field leads to very different magnetometer responses. 

Below, we analyze the NMOR signals versus light power and AC-field amplitude to gain more insight into the difference between sensitivity and bandwidth of the DC and AC magnetometers.

\subsection*{Magnetometer sensitivity}

In the DC magnetometer, the response of the device to the field change is determined by a slope of the in-phase component of the NMOR resonance [Fig.~\ref{fieldchange}(a)]. The strongest response is achieved by maximizing the NMOR-resonance amplitude and minimizing its width. Since both, the resonance amplitude and width, depend on the light power and the AC-field amplitude, optimization of the response is a complex multidimensional procedure.

\begin{figure}[H]
\centering
\includegraphics[width=0.75\linewidth]{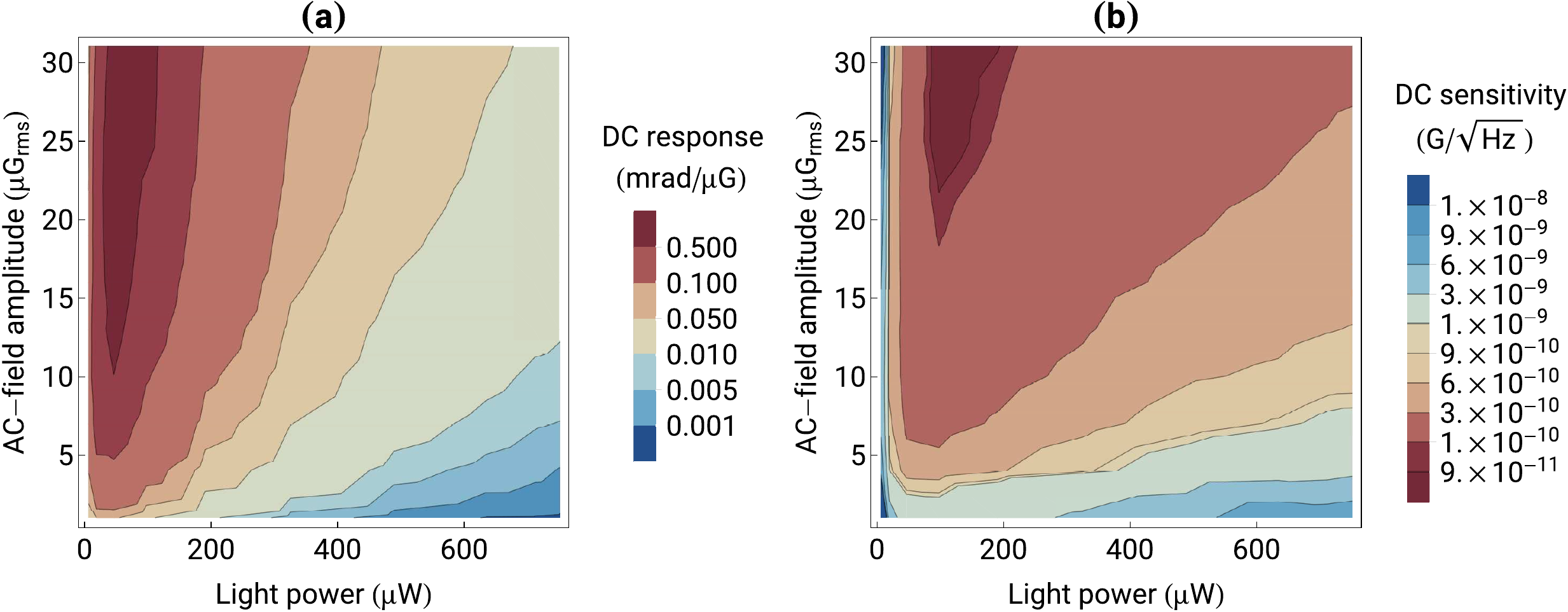}
\caption{(a) Steepness of the central part of the NMOR resonance determining the response of the DC magnetometer to the static-field change and (b) DC sensitivity of the magnetometer as a function of the light power and AC-field amplitude. The sensitivity is determined by projecting the magnetometer response onto the experimentally-determined noise.}
\label{SENSDC}
\end{figure}

To measure the slope of the resonance, the AC-field frequency is scanned across the NMOR resonance (for $\omega_L \approx$ \SI{5.25}{\kilo\hertz}) and the in-phase and quadrature components of the rotation are recorded. The components are respectively fitted with real and imaginary parts of the complex Lorentz function and the amplitude and width of the resonance are determined. The measurement is repeated for different light powers and AC-field amplitudes, which allows us to construct a map of the DC response of the magnetometer [Fig.~\ref{SENSDC}(a)]. 
Data presented in Fig.~\ref{SENSDC}(a) shows that the magnetometer response to the DC-field change monotonically increases with the AC-field amplitude. The increase is nearly linear to about \SI{15}{\micro\gauss}$_{rms}$, when it slowly starts to saturate, but never level out (in an accessible range of parameters). The saturation is manifestation of a transition from linear to nonlinear regime in interaction between light and magnetic field. Particularly, operation in the nonlinear regime, results in broadening of the resonance due to the AC field. The dependence of the magnetometer response on the light power is more complicated. Initially, it increases with the power, reaching its maximum at about \SI{50}{\micro\watt}. Further increase of the power results in deterioration of the response. This effect stems from optical repumping of atoms and destruction of earlier created medium anisotropy. In fact, at higher light powers, the repumping may become a dominant source of relaxation, leading to large power broadening of the resonances. For example, for strong light (\SI{800}{\micro\watt}) the resonance width is nearly 5 times larger than for weak light and associated reduction of the signal amplitude is also observed. This corresponds to a 100-times decrease of the magnetometer response. The analysis of the data presented in Fig.~\ref{SENSDC}(a) reveals that a maximum response of the rotation to the magnetic-field change of about \SI{0.8}{\milli\radian/\micro\gauss} is achieved for light power of about \SI{50}{\micro\watt} and AC-field amplitude of nearly \SI{25}{\micro\gauss}$_{rms}$.

Interestingly, the strongest response of the magnetometer does not translate into the best sensitivity of the device. This is because one needs to take into account noise of the measurements. At the most fundamental level, noise of discussed measurements is determined by quantum nature of photons and atoms (optical shot noise, atomic spin-projection noise, and the back action between light and atoms) (see, for example, Ref.~\cite{grosz2017gawlik} and references therein). On the top of this, there are various other contributions, e.g., those associated with electronics, vibrations, temperature fluctuations, which burden the signal with auxiliary noise, often characterized by the $1/f$ dependence. Finally, there is the environmental magnetic-field noise (uncontrollable magnetic-field fluctuation), which is not inherent to the magnetometer but still affects its performance. For example, in some measurements, environmental noise is orders of magnitude larger than all other contributions severely affecting magnetometer performance. 

To estimate the sensitivity of the DC magnetometer, the photodiode difference signal is recorded without the lock-in amplifier. These measurements are performed for different light powers. In such a way, we not only account for light-intensity independent noise contributions but also include contributions such as photon shot noise. The fast Fourier transform of the polarimeter signal is performed in 1-Hz wide bins in the absence of the AC magnetic field to determine the noise floor. With these measurements, we determine that, close to the nominal resonance frequency $\omega_L$, noise is flat. These measurements also allow to obtain the DC-field sensitivity through normalizing the magnetometer response by the noise [Fig.~\ref{SENSDC}(b)]. 

The analysis of the data shown in Fig.~\ref{SENSDC}(b) reveals that an optimum DC sensitivity of \SI{75}{\pico\gauss\per\sqrt{\hertz}} is achieved for a light power of about \SI{100}{\micro\watt} and an AC-field amplitude of \SI{30}{\micro\gauss}, i.e., parameters different from those optimizing the magnetometer response. Moreover, the DC sensitivity does not deteriorate with the light intensity as quickly as the DC response, shown in Fig.~\ref{SENSDC}(a). This is a consequence of the fact that a signal-to-noise ratio (SNR) of the detected signal (product of the rotation angle and light intensity) improves with the light power, even if polarization rotation starts to slowly deteriorate (amplitude rotation dependence weaker than $1/\sqrt{\textrm{light power}}$), compensating for the rotation-signal deterioration in some range of parameters.

\begin{figure}[H]
\centering
\includegraphics[width=0.75\linewidth]{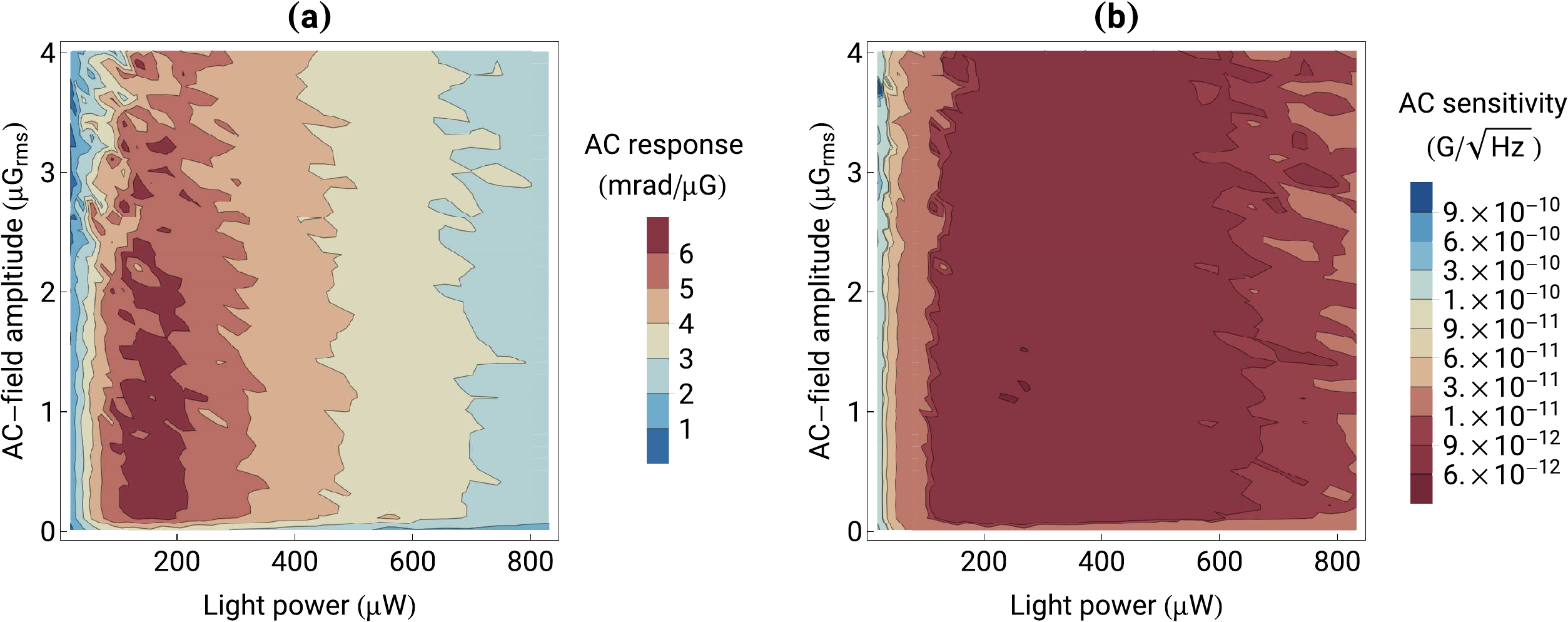}
\caption{(a) Response of the AC magnetometer to the change of the AC-field amplitude and (b) AC-magnetometer sensitivity versus the light power and AC-field amplitude.}
\label{sensAC}
\end{figure}

To determine the response of the NMOR magnetometer to a change of the AC field, the NMOR-resonance amplitude is recorded versus the light power and AC-field amplitude. By taking the difference between NMOR-resonance amplitudes for two successive AC-field values, the response of atoms to the field change is determined. The measurement is then repeated for different light powers and a map, similar to that produced for the DC magnetometer [Fig.~\ref{SENSDC}(a)], is constructed [Fig.~\ref{sensAC}(a)].
Data presented in Fig.~\ref{sensAC}(a) shows that the response of the AC magnetometer to the oscillating-field change is different than that of the DC magnetometer. For instance, the same change in the magnitude of the measured field as in the DC case leads to a response of the AC magnetometer larger by about a factor of 8. Nonetheless, the two magnetometers reveal a similar behavior with respect to the AC-field amplitude; at weak AC fields (up to about \SI{2}{\micro\gauss}$_{rms}$), the response does not depend on $B_{AC}$. However, for stronger fields the sensitivity deteriorates, which is a result of saturation with the oscillating field. Some similarities between the AC and DC responses may be also observed in the light-power dependence. While for lower light powers, the AC response increases with the parameter, it eventually reaches its maximum at about \SI{180}{\micro\watt} and deteriorates for even stronger fields. However, even though the shape of the behavior is similar, its amplitude is very different; in the DC configuration the response of the magnetometer spans 3 orders of magnitude, while in the AC case, the whole change is smaller than a factor of 10 in a whole light-power range. 

The map shown in Fig.~\ref{sensAC}(b) reveals that the sensitivity of the AC magnetometer is roughly and order of magnitude higher than of its DC counterpart. Moreover, the dependence on light power is much weaker, which has important implications (see bandwidth discussion below). The data also shows that the magnetometer reaches an optimum sensitivity of \SI{6}{\pico\gauss\per\sqrt{\hertz}} for a light power of \SI{240}{\micro\watt} and an AC-field amplitude of \SI{1.1}{\micro\gauss}. This is more than 10 times better sensitivity than that obtained with the DC magnetometer under optimal conditions.

In our analysis, the magnetometer responses are determined under the resonance condition ($\omega_{AC}=\omega_L$), when the magnetic-field sensitivity is the strongest. For off-resonant AC fields, when AC-field frequency does not match the Larmor frequency, the responses are weaker. Nonetheless, modification of the AC-field frequency for the DC magnetometer of DC-field magnitude for the AC magnetometer enable tuning of the magnetometers (adjusting the resonance conditions) and maximizing the response. This tunability is one of an important advantages of the NMOR magnetometers. 

\subsection*{Bandwidth considerations}

Apart from the sensitivity, bandwidth is another crucial parameter determining the magnetometer performance. Since a passive DC magnetometer can be considered as an RLC circuit \cite{RLC}, its bandwidth is determined by the width of the observed resonance, given the ground-state relaxation rate. This immediately reveals an opposite dependence of the sensitivity and bandwidth on the resonance width. Thereby, optimization of the magnetometer consists in finding a compromise between how small magnetic field can be detected and how fast such a measurement can be performed. 

To determine the bandwidth of the DC magnetometer, the double demodulation technique is implemented. In this case, a small, sinusoidal, low-frequency ($<$ \SI{200}{\hertz}) modulation of \SI{100}{\nano\gauss}$_{rms}$ is applied to a strong static magnetic field (\SI{7.5}{\milli\gauss}). The amplitude of the modulation is chosen to be large enough to provide a signal with a good SNR, but small enough to still operate within the NMOR resonance. For the measurements, the photodiode difference signal is first demodulated at $\omega_{AC}$ ($\omega_{AC} \approx $ \SI{5.25}{\kilo\hertz}) (the lock-in time constant is small enough not to limit the magnetometer bandwidth), then the in-phase component of the signal is demodulated at the static-field modulation frequency. Next, the frequency is varied and the amplitude of the magnetometer response to the fixed amplitude DC modulation is determined. This allows us to obtain a 3-dB point and hence determine the bandwidth of the magnetometer. After completing a given set of measurements, the procedure is repeated for different light power and AC-field amplitude. This is done until a whole light-power-AC-field-amplitude parameter space is covered and a bandwidth map is produced.

\begin{figure}[H]
\centering
\includegraphics[width=0.375\linewidth]{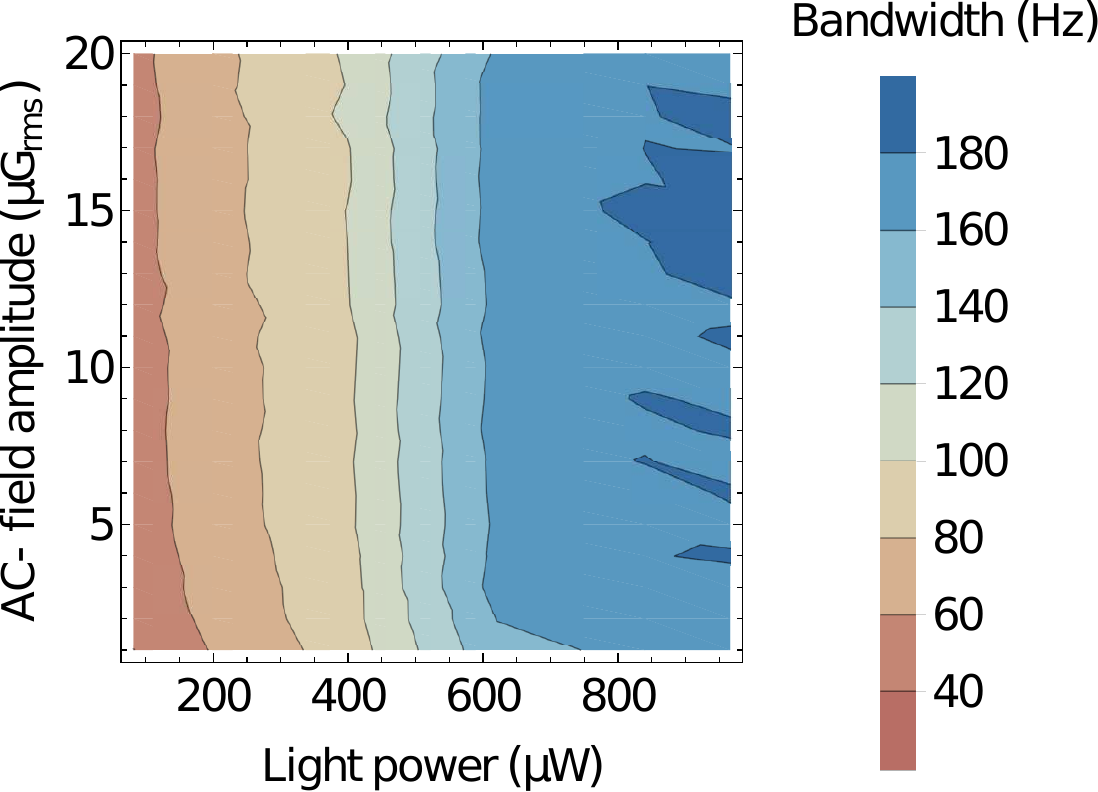}
\caption{DC-magnetometer bandwidth extracted from the 3-dB point using the double-modulation technique. The measurements are performed for different light powers and AC-field amplitudes.}
\label{bandwidth}
\end{figure}

Figure~\ref{bandwidth} shows the bandwidth of the DC magnetometer versus the light power and AC-field amplitude. For more intense light, the ground-state relaxation rate is affected, broadening of the resonance occurs and shortening of the response time of the device (increased bandwidth) is obtained. At the same time, the change of the AC-field amplitude has almost no impact on the resonance width, so that the AC-field just weakly influences the magnetometer bandwidth. This behavior confirms the linear regime of the magnetic-field interaction and nonlinear regime of the optical interaction.

\begin{figure}[H]
\centering
\includegraphics[width=0.75\linewidth]{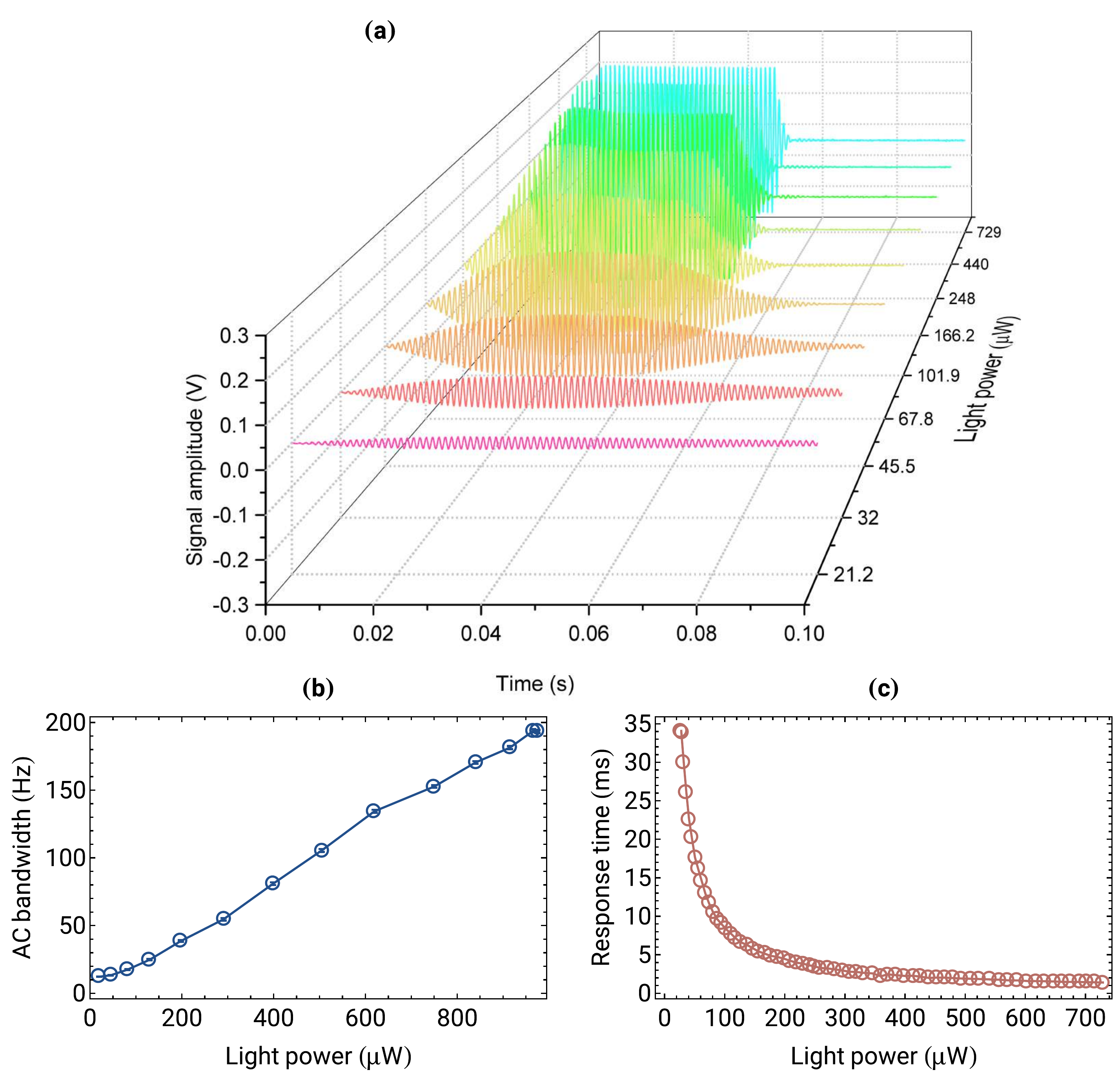}
\caption{(a) Response of the magnetometer to a 50-ms 2-\si{\micro}G$_{rms}$ square pulse of the oscillating magnetic field tuned in resonance with the Larmor frequency. The traces show raw photodiode signals (without demodulation) recorded for different light powers. (b) Bandwidth of the AC magnetometer determined based on the NMOR resonance while scanning the AC-field frequency around the Larmor frequency. Error bars (smaller than measurement points) represent statistical error obtained via fitting complex Lorentzian to the resonance curves. (c) Rise time of the magnetometer determined from the decay of the rotation signal at the end of a pulse (1/$e$ point).}
\label{acbandwidth}
\end{figure}

To investigate the bandwidth of the AC magnetometer, we applied a 50-ms 2-\si{\micro}G$_{rms}$ square pulse of a resonant oscillating magnetic field oriented in the $y$ direction. Figure~\ref{acbandwidth}(a) shows a signal measured with the balanced photodiode (without demodulation). An initially flat magnetometer signal starts to oscillate when the pulse is applied at $t=0$. For low light powers, the response of the device is small and slow, which results in minute and strongly flattened pulse recorded by the magnetometer. The situation changes with increasing light power, when the signal amplitude rises and its square shape is more closely reproduced. The data shows that the bandwidth of the magnetometer increases with the light power [Fig.~\ref{acbandwidth}(b)], reaching hundreds of hertz for the strongest light power. For instance, a 1-ms rising time (1/$e$) of the pulse is achieved for $\approx$ \SI{700}{\micro\watt} [Fig.~\ref{acbandwidth}(c)]. Increasing light power also increases the amplitude of the oscillations. The increase, however, is accompanied by a rise of intensity-dependent noise (not visible in the plots), which leads to some deterioration of the magnetometer measurements [Fig.~\ref{sensAC}(b)].
Operation at higher light power provides several advantages for the AC magnetometry. Particularly, more intense light increases its bandwidth, i.e., rise and fall times of the pulses are significantly reduced. This is beneficial for measuring rapidly decaying magnetic fields (e.g. optical detection of the NMR signal \cite{savukov2007detection}, magnetic particles imaging \cite{dolgovskiy2015quantitative}). At the same time, the strong light decreases the response of the magnetometer to static-field fluctuations (a flattened DC-field dependence), reducing influences of uncontrollable static fields. Eventually, it may lead to the AC-field detection in partially or even completely unshielded environment\cite{wickenbrock2016eddy}.

\section*{Conclusions}

We demonstrated and analyzed the difference in operation of the DC and AC optical magnetometers based on nonlinear magneto-optical rotation. In our case, the measurements of static and oscillating magnetic fields were performed in the same experimental arrangement under very similar experimental conditions. Particular attention was focused on investigations of the sensitivity of the magnetometers. With our results we proved that the resonant character of interaction of atoms and the oscillating field leads to significantly stronger (eight-fold) response of the magnetometer to the AC field than to the DC field. Measurements of magnetometer noise allowed us to study DC- and AC-field sensitivity versus such parameters as light power and AC-field amplitude. With our data we showed initial increase, saturation, and successive deterioration of the sensitivity on the light power and much weaker dependence of the sensitivity on the AC-field amplitude (no deterioration of the signal). These studies allowed us to determined optimum sensitivity of the two magnetometers, which were \SI{75}{\pico\gauss\per\sqrt{\hertz}} for the DC measurements and  \SI{6}{\pico\gauss\per\sqrt{\hertz}} for the AC measurements. This difference is a results of variation in the response of two magnetometers, but also in a signal-to-noise ratio for the optimal light power for AC and DC measurement ($8\sqrt{\SI{240}{\micro\watt}/\SI{100}{\micro\watt}}\approx 12.5$). The DC and AC magnetometer bandwidth was also investigated, revealing roughly linear relation between the bandwidth and the light power. This dependence is clearly visible in the AC-pulse detection, when the magnetometer response at low light power was slow, leading to flattening of the observed pulse. We showed that increasing the power not only improves the bandwidth of the AC magnetometer without significant deterioration of its sensitivity, but also makes the device more immune to the fluctuations of the static fields required for tuning the detection frequency of the magnetometer.

Our results not only contribute to the better understanding of the performance of the DC and AC magnetometers but also suggest approaches toward application of the devices. Particularly, the reduced role of the static magnetic-field fluctuations on the performance of the AC magnetometer, operating within a range up to \SI{1}{\mega\hertz}, opens avenues toward unshielded detection of oscillating fields. This may be particularly interesting in detection of magnetic particles \cite{dolgovskiy2015quantitative}, nuclear magnetic resonance \cite{savukov2007detection}, magnetic resonance imaging in ultra-low fields \cite{xu2006magnetic}, nuclear quadrupole resonance \cite{lee2006subfemtotesla}, and inductive detection of metals \cite{wickenbrock2016eddy,wickenbrockRenzoni,Bevington}.
\section*{Data Availability}
The datasets generated during and/or analyzed during the current study are available from the corresponding author on a reasonable request.
\bibliography{sample}

\begin{thebibliography}{10}
\urlstyle{rm}
\expandafter\ifx\csname url\endcsname\relax
  \def\url#1{\texttt{#1}}\fi
\expandafter\ifx\csname urlprefix\endcsname\relax\def\urlprefix{URL }\fi
\expandafter\ifx\csname doiprefix\endcsname\relax\def\doiprefix{DOI: }\fi
\providecommand{\bibinfo}[2]{#2}
\providecommand{\eprint}[2][]{\url{#2}}

\bibitem{magnetometrybook}
\bibinfo{author}{Budker, D.} \& \bibinfo{author}{Kimball, D.}
\newblock \emph{\bibinfo{title}{Optical Magnetometry}}
  (\bibinfo{publisher}{Cambridge University Press}, \bibinfo{year}{2013}).

\bibitem{dang2010ultrahigh}
\bibinfo{author}{Dang, H.}, \bibinfo{author}{Maloof, A.} \&
  \bibinfo{author}{Romalis, M.}
\newblock \bibinfo{journal}{\bibinfo{title}{Ultrahigh sensitivity magnetic
  field and magnetization measurements with an atomic magnetometer}}.
\newblock {\emph{\JournalTitle{Applied Physics Letters}}}
  \textbf{\bibinfo{volume}{97}}, \bibinfo{pages}{151110}
  (\bibinfo{year}{2010}).

\bibitem{pustelny2013global}
\bibinfo{author}{Pustelny, S.} \emph{et~al.}
\newblock \bibinfo{journal}{\bibinfo{title}{{The Global Network of Optical
  Magnetometers for Exotic physics (GNOME)}: A novel scheme to search for
  physics beyond the standard model}}.
\newblock {\emph{\JournalTitle{Annal. Phys.}}} \textbf{\bibinfo{volume}{525}},
  \bibinfo{pages}{659--670} (\bibinfo{year}{2013}).

\bibitem{safronova2018search}
\bibinfo{author}{Safronova, M.} \emph{et~al.}
\newblock \bibinfo{journal}{\bibinfo{title}{Search for new physics with atoms
  and molecules}}.
\newblock {\emph{\JournalTitle{Reviews of Modern Physics}}}
  \textbf{\bibinfo{volume}{90}}, \bibinfo{pages}{025008}
  (\bibinfo{year}{2018}).

\bibitem{smiciklas2011new}
\bibinfo{author}{Smiciklas, M.}, \bibinfo{author}{Brown, J.},
  \bibinfo{author}{Cheuk, L.}, \bibinfo{author}{Smullin, S.} \&
  \bibinfo{author}{Romalis, M.}
\newblock \bibinfo{journal}{\bibinfo{title}{{New Test of Local Lorentz
  Invariance Using a $^{21}$Ne-Rb-K Comagnetometer}}}.
\newblock {\emph{\JournalTitle{Phys. Rev. Lett.}}}
  \textbf{\bibinfo{volume}{107}}, \bibinfo{pages}{171604}
  (\bibinfo{year}{2011}).

\bibitem{kornack2005test}
\bibinfo{author}{Kornack, T.~W.}
\newblock \emph{\bibinfo{title}{A test of CPT and Lorentz symmetry using a
  {K-3H}e co-magnetometer}}.
\newblock Ph.D. thesis, \bibinfo{school}{Princeton University Princeton, NJ,
  USA} (\bibinfo{year}{2005}).

\bibitem{vasilakis2009limits}
\bibinfo{author}{Vasilakis, G.}, \bibinfo{author}{Brown, J.},
  \bibinfo{author}{Kornack, T.} \& \bibinfo{author}{Romalis, M.}
\newblock \bibinfo{journal}{\bibinfo{title}{{Limits on New Long Range Nuclear
  Spin-Dependent Forces Set with a K-$^3$He Comagnetometer}}}.
\newblock {\emph{\JournalTitle{Physical Review Letters}}}
  \textbf{\bibinfo{volume}{103}}, \bibinfo{pages}{261801}
  (\bibinfo{year}{2009}).

\bibitem{ledbetter2011near}
\bibinfo{author}{Ledbetter, M.} \emph{et~al.}
\newblock \bibinfo{journal}{\bibinfo{title}{Near-zero-field nuclear magnetic
  resonance}}.
\newblock {\emph{\JournalTitle{Physical Review Letters}}}
  \textbf{\bibinfo{volume}{107}}, \bibinfo{pages}{107601}
  (\bibinfo{year}{2011}).

\bibitem{xu2006magnetic}
\bibinfo{author}{Xu, S.} \emph{et~al.}
\newblock \bibinfo{journal}{\bibinfo{title}{Magnetic resonance imaging with an
  optical atomic magnetometer}}.
\newblock {\emph{\JournalTitle{Proceedings of the National Academy of
  Sciences}}} \textbf{\bibinfo{volume}{103}}, \bibinfo{pages}{12668--12671}
  (\bibinfo{year}{2006}).

\bibitem{romalis2011atomic}
\bibinfo{author}{Romalis, M.~V.} \& \bibinfo{author}{Dang, H.~B.}
\newblock \bibinfo{journal}{\bibinfo{title}{Atomic magnetometers for materials
  characterization}}.
\newblock {\emph{\JournalTitle{Materials Today}}}
  \textbf{\bibinfo{volume}{14}}, \bibinfo{pages}{258--262}
  (\bibinfo{year}{2011}).

\bibitem{geomagnetometry}
\bibinfo{author}{Prouty, M.}, \bibinfo{author}{Johnson, R.} \&
  \bibinfo{author}{Hrvoic, A., I.and~Vershovskiy}.
\newblock \bibinfo{title}{{Geophysical Applications}}.
\newblock In \bibinfo{editor}{Budker, D.} \& \bibinfo{editor}{Kimball, D.}
  (eds.) \emph{\bibinfo{booktitle}{Optical Magnetometry}},
  \bibinfo{pages}{329--346} (\bibinfo{publisher}{Cambridge University Press},
  \bibinfo{year}{2013}).

\bibitem{shah2013compact}
\bibinfo{author}{Shah, V.~K.} \& \bibinfo{author}{Wakai, R.~T.}
\newblock \bibinfo{journal}{\bibinfo{title}{A compact, high performance atomic
  magnetometer for biomedical applications}}.
\newblock {\emph{\JournalTitle{Physics in Medicine \& Biology}}}
  \textbf{\bibinfo{volume}{58}}, \bibinfo{pages}{8153} (\bibinfo{year}{2013}).

\bibitem{schultz2016integration}
\bibinfo{author}{Schultz, G.}, \bibinfo{author}{Mhaskar, R.},
  \bibinfo{author}{Prouty, M.} \& \bibinfo{author}{Miller, J.}
\newblock \bibinfo{title}{Integration of micro-fabricated atomic magnetometers
  on military systems}.
\newblock In \emph{\bibinfo{booktitle}{Detection and Sensing of Mines,
  Explosive Objects, and Obscured Targets XXI}}, vol. \bibinfo{volume}{9823},
  \bibinfo{pages}{982318} (\bibinfo{organization}{International Society for
  Optics and Photonics}, \bibinfo{year}{2016}).

\bibitem{higbie2006robust}
\bibinfo{author}{Higbie, J.}, \bibinfo{author}{Corsini, E.} \&
  \bibinfo{author}{Budker, D.}
\newblock \bibinfo{journal}{\bibinfo{title}{Robust, high-speed, all-optical
  atomic magnetometer}}.
\newblock {\emph{\JournalTitle{Review of Scientific Intrum.}}}
  \textbf{\bibinfo{volume}{77}}, \bibinfo{pages}{113106}
  (\bibinfo{year}{2006}).

\bibitem{pustelny2008magnetometry}
\bibinfo{author}{Pustelny, S.} \emph{et~al.}
\newblock \bibinfo{journal}{\bibinfo{title}{Magnetometry based on nonlinear
  magneto-optical rotation with amplitude-modulated light}}.
\newblock {\emph{\JournalTitle{Journal of Applied Physics}}}
  \textbf{\bibinfo{volume}{103}}, \bibinfo{pages}{063108}
  (\bibinfo{year}{2008}).

\bibitem{Chalupczak}
\bibinfo{author}{Chalupczak, W.}, \bibinfo{author}{Godun, R.},
  \bibinfo{author}{Pustelny, S.} \& \bibinfo{author}{Gawlik, W.}
\newblock \bibinfo{journal}{\bibinfo{title}{Room-ftemperature femtotesla
  radio-frequency atomic magnetometer}}.
\newblock {\emph{\JournalTitle{Applied Physics Letters}}}
  \textbf{\bibinfo{volume}{100}}, \bibinfo{pages}{242401}
  (\bibinfo{year}{2012}).

\bibitem{NMOR}
\bibinfo{author}{Budker, D.}, \bibinfo{author}{Kimball, D.},
  \bibinfo{author}{Rochester, S.}, \bibinfo{author}{Yashchuk, V.} \&
  \bibinfo{author}{Zolotorev, M.}
\newblock \bibinfo{journal}{\bibinfo{title}{Sensitive magnetometry based on
  nonlinear magneto-optical rotation}}.
\newblock {\emph{\JournalTitle{Physical Review A}}}
  \textbf{\bibinfo{volume}{62}}, \bibinfo{pages}{043403}
  (\bibinfo{year}{2000}).

\bibitem{ledbetter2007detection}
\bibinfo{author}{Ledbetter, M.} \emph{et~al.}
\newblock \bibinfo{journal}{\bibinfo{title}{Detection of radio-frequency
  magnetic fields using nonlinear magneto-optical rotation}}.
\newblock {\emph{\JournalTitle{Physical Review A}}}
  \textbf{\bibinfo{volume}{75}}, \bibinfo{pages}{023405}
  (\bibinfo{year}{2007}).

\bibitem{Savukov;2005}
\bibinfo{author}{Savukov, I.}, \bibinfo{author}{Seltzer, S.},
  \bibinfo{author}{Romalis, M.} \& \bibinfo{author}{Sauer, K.}
\newblock \bibinfo{journal}{\bibinfo{title}{Tunable atomic magnetometer for
  detection of radio-frequency magnetic fields}}.
\newblock {\emph{\JournalTitle{Physical Review Letters}}}
  \textbf{\bibinfo{volume}{95}}, \bibinfo{pages}{063004}
  (\bibinfo{year}{2005}).

\bibitem{DAVLL}
\bibinfo{author}{Corwin, K.~L.}, \bibinfo{author}{Lu, Z.-T.},
  \bibinfo{author}{Hand, C.~F.}, \bibinfo{author}{Epstein, R.~J.} \&
  \bibinfo{author}{Wieman, C.~E.}
\newblock \bibinfo{journal}{\bibinfo{title}{Frequency-stabilized diode laser
  with the {Z}eeman shift in an atomic vapor}}.
\newblock {\emph{\JournalTitle{Applied Optics}}} \textbf{\bibinfo{volume}{37}},
  \bibinfo{pages}{3295--3298} (\bibinfo{year}{1998}).

\bibitem{grewal2018transient}
\bibinfo{author}{Grewal, R.~S.}, \bibinfo{author}{Pustelny, S.},
  \bibinfo{author}{Rybak, A.} \& \bibinfo{author}{Florkowski, M.}
\newblock \bibinfo{journal}{\bibinfo{title}{Transient dynamics of a nonlinear
  magneto-optical rotation}}.
\newblock {\emph{\JournalTitle{Physical Review A}}}
  \textbf{\bibinfo{volume}{97}}, \bibinfo{pages}{043832}
  (\bibinfo{year}{2018}).

\bibitem{grosz2017gawlik}
\bibinfo{author}{Gawlik, W.} \& \bibinfo{author}{Pustelny, S.}
\newblock \bibinfo{title}{{Nonlinear Magneto-Optical Rotation Magnetometers}}.
\newblock In \bibinfo{editor}{Grosz, A.}, \bibinfo{editor}{Haji-Sheikh, M.~J.}
  \& \bibinfo{editor}{Mukhopadhyay, S.~C.} (eds.)
  \emph{\bibinfo{booktitle}{High sensitivity magnetometers}},
  \bibinfo{pages}{425--450} (\bibinfo{publisher}{Springer},
  \bibinfo{year}{2017}).

\bibitem{RLC}
\bibinfo{author}{W{\l}odarczyk, P.}, \bibinfo{author}{Pustelny, S.},
  \bibinfo{author}{Zachorowski, J.} \& \bibinfo{author}{Lipi{\'n}ski, M.}
\newblock \bibinfo{journal}{\bibinfo{title}{Modeling an optical magnetometer
  with electronic circuits—analysis and optimization}}.
\newblock {\emph{\JournalTitle{Journal of Instrum.}}}
  \textbf{\bibinfo{volume}{7}}, \bibinfo{pages}{P07015} (\bibinfo{year}{2012}).

\bibitem{savukov2007detection}
\bibinfo{author}{Savukov, I.}, \bibinfo{author}{Seltzer, S.} \&
  \bibinfo{author}{Romalis, M.}
\newblock \bibinfo{journal}{\bibinfo{title}{Detection of {NMR} signals with a
  radio-frequency atomic magnetometer}}.
\newblock {\emph{\JournalTitle{Journal of Magnetic Resonance}}}
  \textbf{\bibinfo{volume}{185}}, \bibinfo{pages}{214--220}
  (\bibinfo{year}{2007}).

\bibitem{dolgovskiy2015quantitative}
\bibinfo{author}{Dolgovskiy, V.} \emph{et~al.}
\newblock \bibinfo{journal}{\bibinfo{title}{A quantitative study of particle
  size effects in the magnetorelaxometry of magnetic nanoparticles using atomic
  magnetometry}}.
\newblock {\emph{\JournalTitle{Journal of Magnetism and Magnetic Materials}}}
  \textbf{\bibinfo{volume}{379}}, \bibinfo{pages}{137--150}
  (\bibinfo{year}{2015}).

\bibitem{wickenbrock2016eddy}
\bibinfo{author}{Wickenbrock, A.}, \bibinfo{author}{Leefer, N.},
  \bibinfo{author}{Blanchard, J.~W.} \& \bibinfo{author}{Budker, D.}
\newblock \bibinfo{journal}{\bibinfo{title}{Eddy current imaging with an atomic
  radio-frequency magnetometer}}.
\newblock {\emph{\JournalTitle{Applied Physics Letters}}}
  \textbf{\bibinfo{volume}{108}}, \bibinfo{pages}{183507}
  (\bibinfo{year}{2016}).

\bibitem{lee2006subfemtotesla}
\bibinfo{author}{Lee, S.-K.}, \bibinfo{author}{Sauer, K.},
  \bibinfo{author}{Seltzer, S.}, \bibinfo{author}{Alem, O.} \&
  \bibinfo{author}{Romalis, M.}
\newblock \bibinfo{journal}{\bibinfo{title}{Subfemtotesla radio-frequency
  atomic magnetometer for detection of nuclear quadrupole resonance}}.
\newblock {\emph{\JournalTitle{Applied Physics Letters}}}
  \textbf{\bibinfo{volume}{89}}, \bibinfo{pages}{214106}
  (\bibinfo{year}{2006}).

\bibitem{wickenbrockRenzoni}
\bibinfo{author}{Wickenbrock, A.}, \bibinfo{author}{Tricot, F.} \&
  \bibinfo{author}{Renzoni, F.}
\newblock \bibinfo{journal}{\bibinfo{title}{Magnetic induction measurements
  using an all-optical {87Rb} atomic magnetometer}}.
\newblock {\emph{\JournalTitle{Applied Physics Letters}}}
  \textbf{\bibinfo{volume}{103}}, \bibinfo{pages}{243503}
  (\bibinfo{year}{2013}).

\bibitem{Bevington}
\bibinfo{author}{Bevington, P.}, \bibinfo{author}{Gartman, R.} \&
  \bibinfo{author}{Chalupczak, W.}
\newblock \bibinfo{journal}{\bibinfo{title}{Imaging of material defects with a
  radio-frequency atomic magnetometer}}.
\newblock {\emph{\JournalTitle{Review of Scientific Instruments}}}
  \textbf{\bibinfo{volume}{90}}, \bibinfo{pages}{013103}
  (\bibinfo{year}{2019}).

\end{thebibliography}

\section*{Acknowledgements}
This work was supported by the grant number 2015/19/B/ST2/02129 financed by the Polish National Science Centre.

\section*{Author contributions statement}
K.P. and P.P. conducted the experiment(s), P.P. and S.P. analyzed the results.  All authors contributed to writing and reviewing the manuscript. 

\section*{Additional information}
\textbf{Competing financial interests}\\
The author(s) declare no competing interests.

\end{document}